\documentclass[%
 reprint,
 amsmath,amssymb,
 aps,
]{revtex4-2}

\usepackage{graphicx}
\usepackage{dcolumn}
\usepackage{bm}

\usepackage{lineno,hyperref}
\usepackage{braket}
\usepackage{amsmath,amssymb}
\usepackage{textcomp}

\begin{document}

\preprint{APS/123-QED}

\title{Narrow-line magneto-optical trap for europium}

\author{Yuki Miyazawa}
\author{Ryotaro Inoue}
\author{Hiroki Matsui}
\author{Kenta Takanashi}
\author{Mikio Kozuma}
 \affiliation{
	Department of Physics, Tokyo Institute of Technology, 2-12-1 O-okayama, Meguro-ku, Tokyo 152-8550, Japan}

\date{\today}

\begin{abstract}
We report on the realization of a magneto-optical trap (MOT) for europium atoms using a narrow-line cooling transition with a natural linewidth of 97 kHz. Our starting point is continuous capturing and cooling of optically pumped metastable europium atoms. We have employed simultaneous MOT for the metastable and ground-state atoms. The trapped metastable atoms are successively pumped back to the ground state and then continuously loaded to the narrow-line MOT, where up to $4.7\times10^7$ atoms are captured. A spin-polarized sample at a temperature of $6\,\mathrm{\mu K}$ and with a peak number density of $2.2\times10^{11}\,\mathrm{cm^{-3}}$ is obtained through the compression process, resulting in a phase space density of $3\times10^{-5}$.

\end{abstract}

\maketitle


\section{Introduction}
Quantum degenerate gas of atoms with a large dipole moment has opened up a new avenue for studying ultracold atoms with long-range and anisotropic interaction \cite{Lahaye2008, Chomaz2019, Baier201}. These quantum degenerate gases were first realized by using chromium \cite{Griesmaier2005} with a magnetic moment of $6\,\mu_B$; subsequently, the dipolar lanthanide atoms of dysprosium \cite{Lu2011} and erbium \cite{Aikawa2012} with a magnetic moment of $10\,\mu_B$ and $7\,\mu_B$, respectively, too were used. Europium (Eu) also belongs to the lanthanide group and has a large dipole moment of $7\,\mu_B$. It has two stable bosonic isotopes: ${}^{151}$Eu and ${}^{153}$Eu with natural abundances of $48\,\%$ and $52\,\%$, respectively. Since both isotopes have nuclear spins of $I = 5/2$, they have hyperfine structures in their ground state. This enables us to control their scattering length using radiofrequency fields \cite{Hanna_2010, Papoular2010}: such control is useful for examining the behavior of dipolar spinor Bose gases under an ultralow magnetic field \cite{Kawaguchi2006, Kawaguchi2006a, Takahashi2007}. Since the electronic angular momentum of the ground state is zero, a less dense and nonchaotic Feshbach resonance spectrum is expected in the case of Eu, which is in contrast to the spectra of other dipolar lanthanides \cite{ZarembaKopczyk2018, Maier2015, Khlebnikov2019}.

The laser cooling of other lanthanide atoms was successfully conducted by combining Zeeman slowing using a broad optical transition and a magneto-optical trap (MOT) with a narrow optical transition \cite{Frisch2012, Maier:14, Sukachev2014}. Cold atoms in the MOT were loaded to an optical dipole trap, and quantum degenerate gases were obtained through successive evaporative cooling \cite{Lu2011, Aikawa2012, Davletov2020}. However, this approach cannot be directly applied for Eu atoms because the optical transition broad enough for Zeeman slowing has large optical leaks; excited atoms decay to at least six metastable states with a total probability of $1.05(2)\times10^{-3}$ \cite{MIYAZAWA2017171}. Hence, we pumped atoms to another metastable state that exhibits a quasicyclic transition at 583 nm and implemented the Zeeman slowing and MOT \cite{Inoue2018}. In this paper, we report on the laser cooling of Eu in the ground state. We optically pumped the metastable atoms back to the ground state and implemented a narrow-line MOT such that the MOT operations for the metastable and ground states of atoms were simultaneously performed.

The rest of the paper is organized as follows. In section \ref{sec:2}, the narrow-line MOT procedure is described with focus on optical transitions. Our experimental setup is explained in section \ref{sec:3}. The experimental results of MOT characteristics are presented in section \ref{sec:4}, and the paper is concluded in section \ref{sec:5}.

\section{\label{sec:2}Narrow-line MOT procedure}
\begin{figure*}[htb]
	\includegraphics{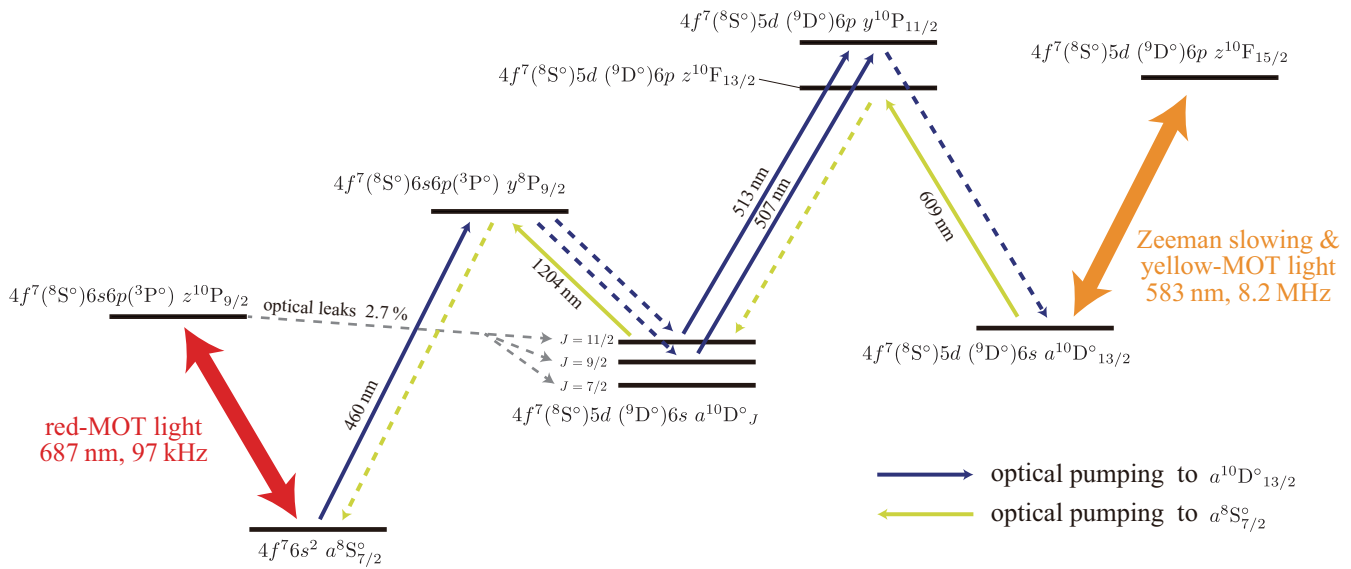}
	\caption{\label{fig:1}Energy levels and transitions of Eu relevant for our laser cooling procedure \cite{Fechner1987, Hartog2002}. Solid lines indicate laser-driven transitions and dashed lines indicate spontaneous decay channels.} 
\end{figure*}

Figure \ref{fig:1} shows the schematic of our laser cooling procedure. We start from a hot atomic beam of Eu in the ground state. Atoms in the ground state $a^8{\rm S}_{7/2}^\circ$ are transferred to the metastable state $a^{10}{\rm D}_{13/2}^\circ$ as described below. By driving the $460\,\mathrm{nm}$ transition, atoms are pumped from the ground state into two intermediate states $a^{10}{\rm D}_{9/2}^\circ$ and $a^{10}{\rm D}_{11/2}^\circ$; then, they are pumped to $a^{10}{\rm D}_{13/2}^\circ$ by driving the $507\,\mathrm{nm}$ and $513\,\mathrm{nm}$ transitions, respectively. The total transfer efficiency is estimated as up to $19\,\%$ \cite{MIYAZAWA2017171, Hartog2002}. The atoms in $a^{10}{\rm D}_{13/2}^\circ$ are then Zeeman slowed and captured in a MOT (yellow-MOT) using the $a^{10}{\rm D}_{13/2}^\circ \leftrightarrow z^{10}{\rm F}_{15/2}$ cooling transition at $583\,\mathrm{nm}$ with a natural linewidth of $\Gamma_{583}/2\pi = 8.2\,\mathrm{MHz}$ \cite{Hartog2002}. The captured metastable atoms are successively pumped back to the ground state with the pumping-back efficiency of $94\,\%$ \cite{MIYAZAWA2017171, Hartog2002} by driving two transitions at the wavelengths of $609\,\mathrm{nm}$ and $1204\,\mathrm{nm}$. Pumped back atoms are cooled and trapped in the narrow-line MOT (red-MOT) using the $a^{8}{\rm S}_{7/2}^\circ \leftrightarrow z^{10}{\rm P}_{9/2}$ cooling transition at $687\,\mathrm{nm}$ with a natural linewidth of $\Gamma_{687}/2\pi  = 97\,\mathrm{kHz}$ \cite{Fechner1987}. Note that this series of optical pumping and laser cooling processes were executed simultaneously. Atoms in the ground state are thus loaded to the red-MOT continuously. The cooling transition at $687\,\mathrm{nm}$ has optical leaks from the excited state $z^{10}{\rm P}_{9/2}$ to three metastable states $a^{10}{\rm D}_{7/2}^\circ$, $a^{10}{\rm D}_{9/2}^\circ$, and $a^{10}{\rm D}_{11/2}^\circ$. The decay rates were experimentally measured in this work (see Appendix \ref{app:B}), and the sum of them was estimated as $1.64(5)\times 10^4\, \mathrm{s^{-1}}$, which corresponds to a branching fraction of  $2.7\, \%$. In this work, we optically repump the atoms from the metastable states back to the ground state via $y^{8}{\rm P}_{9/2}$ state (see Appendix \ref{app:A}).

\section{\label{sec:3}Experimental setup}

Our Eu atomic beam was produced from an effusive oven operating at $770\, \mathrm{K}$. Optically pumped metastable atoms were Zeeman slowed and loaded to the yellow-MOT, which was formed in a quadrupole magnetic field provided by anti-Helmholtz coils and three pairs of counter-propagating, circularly polarized cooling-light beams at $583\,\mathrm{nm}$. The beam diameter was about $25\, \mathrm{mm}$ and truncated by a circular aperture of $21.6\, \mathrm{mm}$ diameter. Optical pumping to the ground state was performed by two-color lights at $609\, \mathrm{nm}$ and $1204\, \mathrm{nm}$. The beam diameter of the pumping light at $609\, \mathrm{nm}$ matched the size of the yellow-MOT of about $500\, \mu \mathrm{m}$ so that cooled atoms were selectively pumped back.

The narrow-line red-MOT was formed in the same magnetic field as the yellow-MOT. The cooling light was produced by an external cavity laser diode and amplified by a tapered amplifier. The laser linewidth was suppressed below $33\, \mathrm{kHz}$ by stabilizing the laser frequency to the resonance of an ultralow-expansion cavity. The cooling laser beams overlapped the $583\mathchar`- \mathrm{nm}$ beams with the same polarization and beam diameter. The  intensity ratio of the laser beams propagating in axial and radial directions of the coil was set as 2:1. Here, the axial direction was parallel to gravity.  The cooling light was red-detuned by $\delta_{687}$ with respect to the $F=6\leftrightarrow F'=7$ cyclic transition. We have added some additional lasers to repump from the levels $a^{10}\mathrm{D}$ (see Appendix \ref{app:A}  for details).

The number of trapped atoms was determined by an absorption imaging technique using the $460\,\mathrm{nm}$ transition with spin polarization. We turned off all the lights and quadrupole magnetic field after loading the atoms and applied a magnetic field of approximately $10^{-4}\, \mathrm{T}$ along the probe axis. Then, atoms were optically pumped to the $\ket{F=6, m_F =6}$ Zeeman sublevel by a two-color $\sigma_+$ polarized light pulse tuned to $F=5\leftrightarrow F'=6$ and $F=6\leftrightarrow F'=6$ transitions. Subsequently, the polarized atoms were illuminated by a $\sigma^+$ polarized probe laser beam near-resonant on $\ket{F=6, m_F=6}\leftrightarrow\ket{F'=7, m_{F'}=7}$ transition. The absorption by the atoms casts a shadow on an imaging device. We confirmed that the degree of the spin polarization is sufficient for determining the number of atoms within a few percent errors by comparing the optical depth of two images obtained by $\sigma_+$ and $\sigma_-$ polarized probe beams.

\section{\label{sec:4}Experimental results}

\begin{figure}[htb]
		\includegraphics{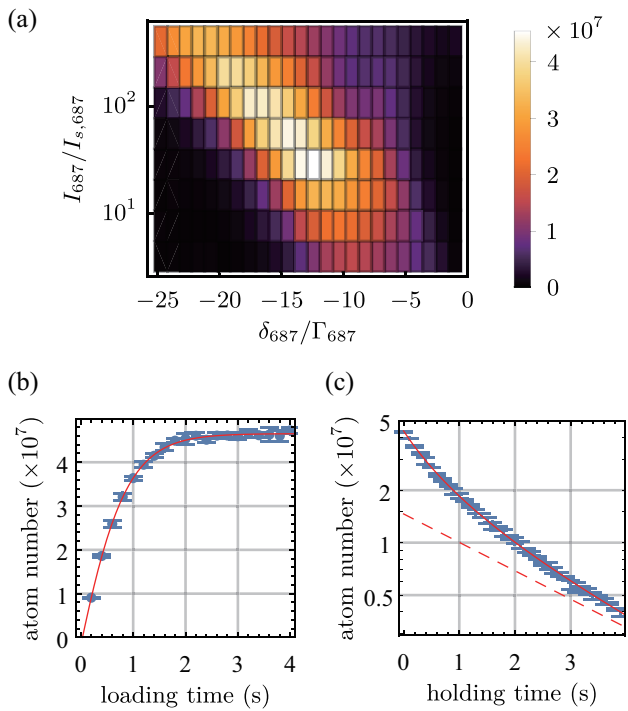}
		\caption{\label{fig:2}(a) The number of trapped atoms as a function of the detuning $\delta_{687}$ and the intensity $I_{687}$. The parameters for the yellow-MOT are fixed to the optimal conditions (see text). Loading (b) and decay (c) curves of the red-MOT under the optimal conditions. The solid lines are fitted to the data, and the dashed line in (c) is the exponential asymptote of the fit.} 
\end{figure}

By using the aforementioned laser-cooling procedure, up to $4.7\times 10^7$ ${}^{151}$Eu atoms were trapped. The number of atoms in our red-MOT was maximized under the following experimental conditions. The axial magnetic field gradient was set to $\partial B_z/\partial z=3\times 10^{-2}\, \mathrm{T/m}$. The detuning and total intensity of the cooling laser for the yellow-MOT were $-1.75 \, \Gamma_{583}$ and $3.7\, I_{s, 583}$, respectively, whereas these values for the narrow-line red-MOT were $\delta_{687}=-13\, \Gamma_{687}$ and $I_{687}=40\, I_{s, 687}$. Here, $I_{s, 583}=5.4\, \mathrm{mW/cm^2}$, and $I_{s, 687}=3.9\, \mu \mathrm{W/cm^2}$ were the saturation intensities for the cooling transitions. Figure \ref{fig:2} (a) summarizes the number of trapped atoms after $4\, \mathrm{s}$ of loading as a function of the detuning $\delta_{687}$ and the intensity $I_{687}$. The maximum number of atoms is obtained around $\delta_{687} = -13\, \Gamma_{687}$ and $I_{687} = 40 \, I_{s, 687}$. The loading and decay curves of the red-MOT under the optimal conditions are shown in Fig.~\ref{fig:2} (b) and (c), respectively. From curve fitting \cite{Browaeys2000},  we obtained a loading rate $R=6\times 10^7\, \mathrm{atoms/s}$, a one-body loss rate of $1/\tau =0.38\, \mathrm{s^{-1}}$, and a two-body loss rate of $\beta =7\times 10^{-11}\, \mathrm{cm^3/s}$. The measured one-body loss rate is well described by the incompleteness of our repumping scheme (see Appendix \ref{app:A}) assuming that the population in the excited state is 0.02. This population is estimated from equilibrium of the radiative force, gravity, and magnetic force \cite{Dreon2017}. Note that our vacuum chamber with $<10^{-9}\,\mathrm{Pa}$ background pressure does not limit the decay rate.

\begin{figure}[htb]
		\includegraphics{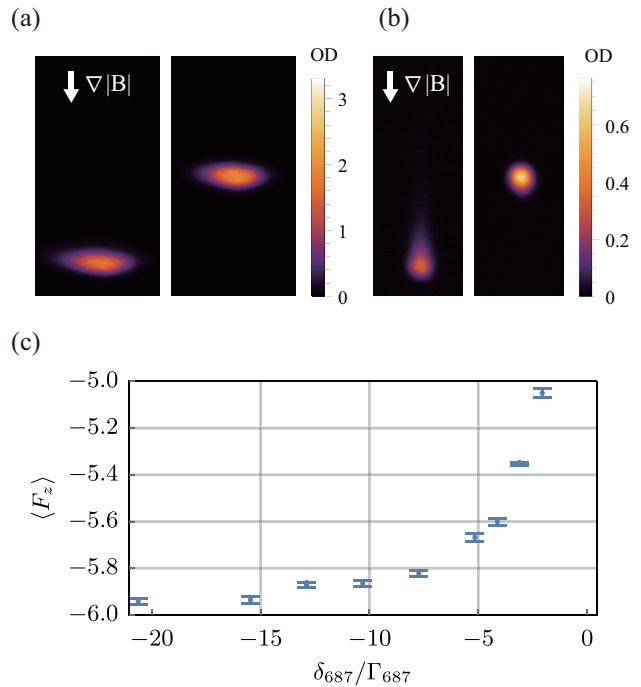}
		\caption{\label{fig:3}Spontaneous spin polarization in the narrow-line red-MOT.  (a and b) Absorption images are taken at the same detuning of  $\delta _{687}=-21\,\mathrm{\Gamma_{687}}$ (a) and $\delta _{687}=-2\,\mathrm{\Gamma_{687}}$ (b). The left images are taken after free expansion with applying a magnetic field gradient for spin separation, whereas the right images are taken in the absence of the magnetic field gradient. (c) The estimated mean spin projection $\langle F_z \rangle$ as a function of the detuning $\delta_{687}$. The intensity is  $I_{687}=3\, I_{s, 687}$, and the axial magnetic field gradient is $\partial B_z/\partial z=3\times 10^{-2}\, \mathrm{T/m}$. } 
\end{figure}

One of the features of a narrow-line MOT is the spontaneous spin polarization of the atoms in the MOT, as observed in the narrow-line MOTs of other lanthanide atoms \cite{Aikawa2012, Dreon2017}. As shown in Fig.~\ref{fig:3}, we evaluated the mean spin projection of the atoms in our red-MOT by using the Stern--Gerlach effect. After releasing the atoms from the red-MOT, we applied a 7-ms-long vertical magnetic field gradient of about $0.3\, \mathrm{T/m}$. An absorption image was then obtained in the same manner as explained in section \ref{sec:2}. The typical absorption images are shown in Fig.~\ref{fig:3}  (a) and (b) along with the reference images, which were obtained without applying a magnetic field gradient. The populations of individual spin levels were estimated with multiple Gaussian fitting of the absorption images, considering the corresponding reference images. We estimated the population distributions as a function of the detuning $\delta_{687}$, and calculated the mean spin projections $\braket{F_z}=\sum_{m_F=-6}^{6}p_{m_F} m_F$ from the obtained populations $p_{m_F}$ as shown in Fig.~\ref{fig:3}(c). The figure reveals that atoms were well spin-polarized for larger detuning. With the detuning of $\delta_{687}=-21\, \Gamma_{687}$, $95\, \%$ of the atoms populated the lowest Zeeman sublevel $\ket{F=6,\,m_F=-6}$.

To increase the phase space density of the atomic cloud, a compression sequence was performed after the loading stage. Using the optimal parameters $\partial B_z/\partial z=2.5\times 10^{-2}\, \mathrm{T/m}$, $I_{687}=2\, I_{s, 687}$, and $\delta_{687}=-6\,  \Gamma_{687}$, we achieve a temperature of $6\, \mu \mathrm{K}$ and a peak number density of $2.2\times 10^{11}\, \mathrm{atoms/cm^3}$ with an atom number of $3.3\times 10^7$. Under this condition, the population of $\ket{F=6,\,m_F=-6}$ was inferred as about $0.8$ from our Stern--Gerlach experiment, indicating a phase-space density of about $3\times 10^{-5}$.

\section{\label{sec:5}Conclusion}
In conclusion, we demonstrated a MOT for Eu atoms using the narrow-line transition with a natural linewidth of $97\, \mathrm{kHz}$. Although Eu has no suitable transition for Zeeman slowing of an atomic beam in the ground state, atoms are successfully loaded to the narrow-line MOT by using Zeeman slowing and a MOT operated at the 583-nm cooling transition originating from the metastable state in a continuous manner. The phase space density and the number of atoms after the compression stage provide us with good starting conditions for direct loading to an optical dipole trap.

\section*{Acknowledgments}
This study was supported by JSPS KAKENHI Grants Numbers JP16K13856 and JP17J06179. Y.M. acknowledges partial support from the Japan Society for the Promotion of Science.

\appendix
\section{\label{app:A}Repumping scheme}
\begin{figure}[htb]	
	\includegraphics{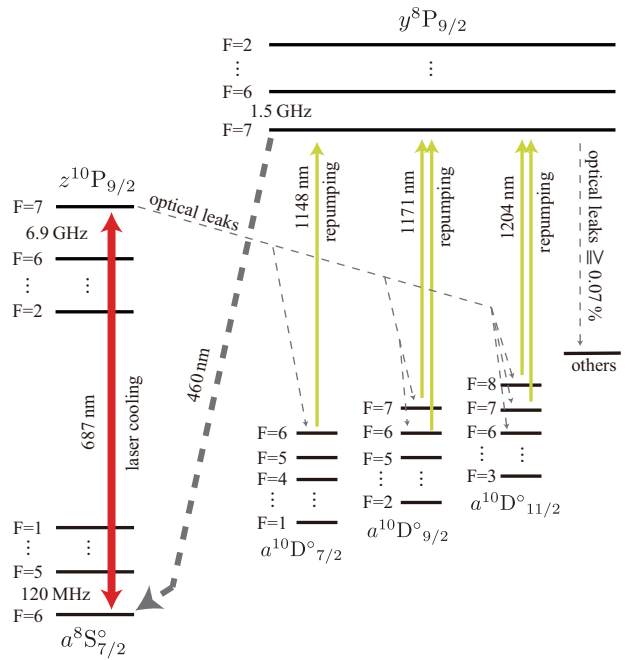}
	\caption{\label{fig:4}Energy levels and transitions of ${}^{151}$Eu relevant for our repumping scheme including hyperfine structures. Solid lines indicate laser-driven transitions and dashed lines indicate spontaneous decay channels. Hyperfine splittings are calculated based on hyperfine constants in references \cite{Zaal1979, Jin2002}.} 
\end{figure}

The 687-nm cooling transition has optical leaks to some metastable states. The upper-state $\ket{z^{10}{\rm P}_{9/2}, F=7}$ has six dominant decay channels to $\ket{a^{10}{\rm D}_{7/2}^\circ, F=6}$, $\ket{a^{10}{\rm D}_{9/2}^\circ, F=6,7}$, and $\ket{a^{10}{\rm D}_{11/2}^\circ, F=6, 7, 8}$, besides one to the lower-state hyperfine manifold. In this work, we plugged merely the five most dominant relaxation pathways as schematically shown in Fig.~\ref{fig:4}. Here we omit the repumping from the state $\ket{a^{10}{\rm D}_{11/2}^\circ, F=6}$ since the optical leak probability to this state is comparable with that to the other states referred to as ``others'' in Fig.~\ref{fig:4}. One can estimate this overall repumping efficiency of the scheme as up to $99.9\,\%$ utilizing some transition probabilities shown in Table \ref{tab:1} and \ref{tab:2}, assuming that all five repumpers have sufficient intensities. In our experiment, each repumping light beam was introduced to the MOT chamber and was retro-reflected, where the beam diameter and the power were set as $3\,\mathrm{mm}$ and $\sim10\,\mathrm{mW}$, respectively.

\begin{table}[htb]
	\caption{\label{tab:1}Transition wavelengths $\lambda$ and transition probabilities $A_{ki}$ in our repumping scheme.}
	\begin{ruledtabular}
	\begin{tabular}{cccc}		
		\begin{tabular}{c} upper\\ level  \end{tabular}  &  \begin{tabular}{c} lower\\ level  \end{tabular} & $\lambda\ (\mathrm{nm})$  &  $A_{ki}\ (\mathrm{s^{-1})}$ \\ \hline
		$y^8{\rm P}_{9/2}$ & $a^8{\rm S}_{7/2}^{\circ}$ & 460 & $1.61(8)\times 10^8$ \cite{Hartog2002}  \\ 
		& $a^{10}{\rm D}_{7/2}^{\circ}$ & 1148  &  $0.5(1)\times 10^4$ \cite{MIYAZAWA2017171}\\
		& $a^{10}{\rm D}_{9/2}^{\circ}$ & 1171  & $1.7(1)\times 10^4$ \cite{MIYAZAWA2017171} \\
		& $a^{10}{\rm D}_{11/2}^{\circ}$ & 1204  & $2.9(2)\times 10^4$ \cite{MIYAZAWA2017171} \\
		& others & 1577-4880 & $\geqq11.7(6)\times 10^4$ \cite{MIYAZAWA2017171} \\
	\end{tabular}
	\end{ruledtabular}

\end{table}

\section{\label{app:B}Transition probabilities for optical-leak lines}
We measured the transition probabilities for the leakage transitions from $z^{10}{\rm P}_{9/2}$ to $a^{10}{\rm D}_{7/2}^\circ$, $a^{10}{\rm D}_{9/2}^\circ$, and $a^{10}{\rm D}_{11/2}^\circ$. After loading the atoms, we turned off one of the three repumping beams, and kept an optical leak unplugged. Under this condition, the lifetime of the red-MOT is limited by the optical leak. By measuring the lifetime of the red-MOT, we estimated the transition probability for the unplugged leakage transition. In this manner, we measured the three transition probabilities, which are listed in Table \ref{tab:2}.

\begin{table}[htb]
	\caption{\label{tab:2}The transition wavelengths $\lambda$ and measured transition probabilities $A_{ki}$ for the leakage transitions.}
	\begin{ruledtabular}
	\begin{tabular}{cccc}
		\begin{tabular}{cc} upper\\ level  \end{tabular}  &  \begin{tabular}{cc} lower\\ level  \end{tabular} & $\lambda\ (\mathrm{nm})$  &  $A_{ki}\ (\mathrm{s^{-1})}$ \\ \hline	
		$z^{10}{\rm P}_{9/2}$ 
		& $a^{10}{\rm D}_{7/2}^{\circ}$ & 6602 & $4.5(2)\times 10^3$ \\
		& $a^{10}{\rm D}_{9/2}^{\circ}$ & 7454 & $7.4(2)\times 10^3$ \\
		& $a^{10}{\rm D}_{11/2}^{\circ}$ & 9039 & $4.5(1)\times 10^3$ \\
	\end{tabular}
	\end{ruledtabular}
\end{table}

\bibliographystyle{apsrev4-2}
\bibliography{mybibfile}

\end{document}